\documentclass[3p, 12pt, sort&compress,fleqn]{elsarticle}

\usepackage{amsmath,amssymb,graphicx,siunitx,float,mathtools}
\usepackage[table]{xcolor}
\usepackage{hyperref}
\hypersetup{colorlinks=true,breaklinks=true,allcolors=blue}

\newcommand\revision[1]{#1}

\usepackage{geometry}
\makeatletter
\def\ttabular{%
\hbox\bgroup
\let\\\cr
\def\rulea{\ifnum\rowc=\@ne \hrule height 1.3pt \fi}
\def\ruleb{
\ifnum\rowc=1\hrule height 1.3pt \else
\ifnum\rowc=6\hrule height \heavyrulewidth 
   \else \hrule height \lightrulewidth\fi\fi}
\valign\bgroup
\global\rowc\@ne
\rulea
\hbox to 10em{\strut \hfill##\hfill}%
\ruleb
&&%
\global\advance\rowc\@ne
\hbox to 10em{\strut\hfill##\hfill}%
\ruleb
\cr}
\def\endttabular{%
\crcr\egroup\egroup}

\allowdisplaybreaks[4]

\makeatletter
\def\ps@pprintTitle{%
 \let\@oddhead\@empty
 \let\@evenhead\@empty
 \def\@oddfoot{}%
 \let\@evenfoot\@oddfoot}
\makeatother

\begin{document}

\begin{frontmatter}

\title{Approximate analytical solution for transient heat and mass transfer\\ across an irregular interface}

\author[qut]{Elliot J. Carr}
\ead{elliot.carr@qut.edu.au}
\author[qut]{Dylan J. Oliver}
\author[qut]{Matthew J. Simpson}
\address[qut]{School of Mathematical Sciences, Queensland University of Technology (QUT), Brisbane, Australia.}


\begin{abstract}
Motivated by practical applications in heat conduction and contaminant transport, we consider heat and mass diffusion across a perturbed interface separating two finite regions of distinct diffusivity. Under the assumption of continuity of the solution and diffusive flux at the interface, we use perturbation theory to develop an asymptotic expansion of the solution valid for small perturbations. Each term in the asymptotic expansion satisfies an initial-boundary value problem on the unperturbed domain subject to interface conditions depending on the previously determined terms in the asymptotic expansion. Demonstration of the perturbation solution is carried out for a specific, practically-relevant set of initial and boundary conditions with semi-analytical solutions of the initial-boundary value problems developed using standard Laplace transform and eigenfunction expansion techniques. Results for several choices of the perturbed interface confirm the perturbation solution is in good agreement with a standard numerical solution.
\end{abstract}

\begin{keyword}
diffusion, perturbation, interface, semi-analytical.
\end{keyword}

\end{frontmatter}

\section{Introduction}
\noindent The diffusion equation is fundamental to applied mathematics with numerous applications in engineering and the physical and life sciences \cite{liu_1998,pontrelli_2020,simpson_2017,zhao_2016,mantzavinos_2016}. Most practical applications of the diffusion equation involve a complex heterogeneous geometry of irregular shape exhibiting spatial variation in diffusivity. While computing numerical solutions to such problems is relatively straightforward, analytical solutions remain highly sought after since they provide greater insight into the physical significance of key model parameters \cite{ozisik_1993}, generally exhibit higher accuracy and can be evaluated at any point in continuous space and time. Of most interest to this work are analytical solutions for the steady-state diffusion equation in irregular geometries such as circles, ellipses or rectangles with perturbed boundaries and constant diffusivity \cite{simpson_2021,scheffler_1974,jiji_2009,aziz_1980} and analytical solutions of the transient diffusion equation in layered media with parallel/concentric interfaces between regions of distinct diffusivity~ \cite{carr_2016,kaoui_2018,rodrigo_2016,hickson_2009,ozisik_1993,carr_2020c}. 

\begin{figure}[h]
\centering
\includegraphics[width=0.9\textwidth]{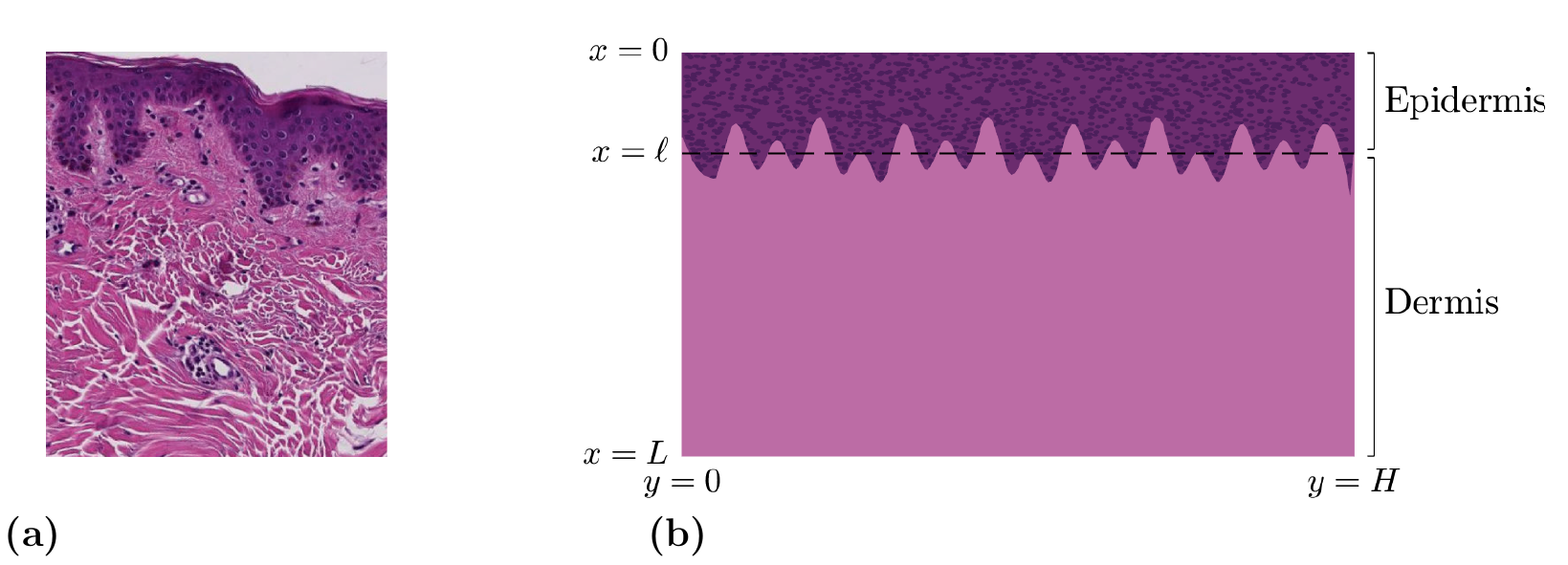}
\caption{\textbf{Heat conduction in heterogeneous skin}. (a) Haemotoxylin and eosin (H\&E) staining of real skin showing a well-defined irregular interface separating the epidermis (purple) and dermis (pink) (adapted from \cite{haridas_2017}). (b) Schematic diagram of a heterogeneous skin medium of depth $L$ and width $H$ with an irregular interface described by $x = \ell + \varepsilon w(y)$. We solve the heat equation across the medium, with distinct diffusivities in the epidermis (purple) and dermis (pink) regions and continuity of temperature and diffusive flux imposed at $x = \ell + \varepsilon w(y)$ \cite{mcinerney_2019}.}
\label{fig:skin}
\end{figure}

In this paper, we consider the transient diffusion equation in a two-dimensional heterogeneous medium comprising two regions separated by an irregular interface. \revision{Our interest in this problem is mainly motivated by the work of \citet{mcinerney_2019} who studied heat conduction in living heterogeneous skin as a way to better understand scald burn treatments. Other motivations include the work of \citet{carr_2019} who analysed heat conduction in two-layer solid materials to develop formulae for calculating the thermal diffusivity of the two constituent materials and \citet{chen_2009} who considered groundwater contamination by way of diffusion of contaminants through manufactured two-layer composite landfill liners. In each of these articles, the authors assume the two layers are separated by a (perfectly) horizontal or vertical interface, which gives rise to a simplified one-dimensional two-layer mathematical model governing the spatial and temporal dynamics of the solution (temperature or concentration in the above applications). 

In practice, interfaces are seldom (perfectly) vertical or horizontal but irregular, as demonstrated for skin in Figure \ref{fig:skin}. In this case, diffusive transport can be described by the following set of equations:
\begin{gather}
\label{eq:pde1}
\frac{\partial u_{1}}{\partial t} = D_{1}\Delta u_{1},\quad 0 < x < \ell + \varepsilon w(y),\enspace 0 < y < H,\enspace t > 0,\\
\label{eq:pde2}
\frac{\partial u_{2}}{\partial t} = D_{2}\Delta u_{2},\quad \ell + \varepsilon w(y) < x < L,\enspace 0 < y < H,\enspace t > 0,\\
\label{eq:int1}
u_{1}(\ell + \varepsilon w(y),y,t) = u_{2}(\ell + \varepsilon w(y),y,t),\quad 0 < y < H,\enspace t > 0,\\
\label{eq:int2}
D_{1}\nabla u_{1}(\ell + \varepsilon w(y),y,t)\cdot\mathbf{n}(y) = D_{2}\nabla u_{2}(\ell + \varepsilon w(y),y,t)\cdot\mathbf{n}(y),\quad 0 < y < H,\enspace t > 0,
\end{gather}
where $u_{1}$ and $u_{2}$ denote the temperature in the first and second regions; $x = \ell + \varepsilon w(y)$ specifies the interface separating the two regions; $\mathbf{n}(y)\in\mathbb{R}^{2}$ denotes a vector normal to $x = \ell + \varepsilon w(y)$; and $D_{1}$ and $D_{2}$ denote the distinct diffusivity values in the first and second regions. Continuity of both temperature and diffusive flux is imposed at the interface by way of conditions (\ref{eq:int1}) and (\ref{eq:int2}).} 

In this paper, we present an approximate analytical solution of equations (\ref{eq:pde1})--(\ref{eq:int2}), subject to appropriate initial and boundary conditions, by developing asymptotic expansion of the solutions $u_{1}$ and $u_{2}$ valid for small non-negative $\varepsilon$. Using perturbation theory, we show that each term in the asymptotic expansion satisfies an initial-boundary value problem on the unperturbed two-layer domain (i.e., $\varepsilon = w(y) = 0$) with interface conditions depending on the previously computed terms in the expansion. Each initial-boundary value problems is then solved semi-analytically using the Laplace transform for a specific, practically-relevant set of initial and boundary conditions representing scalding at the surface, $x = 0$ \cite{mcinerney_2019}. Resulting solution fields are visualised for several different choices for the perturbed interface and verified against numerical solutions obtained via a standard finite volume discretisation. MATLAB code implementing our perturbation solution and allowing the solution approach to be investigated for other appropriate choices of the parameters ($\varepsilon$, $w(y)$, $\ell$, $D_{1}$, $D_{2}$, $L$ and $H$) is made available on a GitHub repository: \href{https://github.com/elliotcarr/Carr2022a}{https://github.com/elliotcarr/Carr2022a}.

\section{Perturbation solution}
\label{sec:solution_method}
\noindent We assume the solution of equations (\ref{eq:pde1})--(\ref{eq:int2}), subject to appropriate initial and boundary conditions, can be expanded in powers of $\varepsilon$: 
\begin{align}
\label{eq:ansatz1}
u_{1}(x,y,t) = \sum_{i=0}^{\infty} \varepsilon^{i}u_{1}^{(i)}(x,y,t),\\
\label{eq:ansatz2}
u_{2}(x,y,t) = \sum_{i=0}^{\infty} \varepsilon^{i}u_{2}^{(i)}(x,y,t),
\end{align}
where the terms $u_{1}^{(i)}$ and $u_{2}^{(i)}$ for $i\in\mathbb{N} := \{0,1,2,\hdots\}$ satisfy the transient diffusion equation on the unperturbed domain, that is, equation (\ref{eq:pde1}) on $0 < x < \ell$ and equation (\ref{eq:pde2}) on $\ell < x < L$ \cite{simpson_2021,scheffler_1974}.

Now consider the interface condition (\ref{eq:int1}). Expanding both sides of (\ref{eq:int1}) in a Taylor series centered at $x = \ell$,
\begin{align*}
\sum_{k=0}^{\infty} \frac{\varepsilon^{k}w(y)^{k}}{k!} \frac{\partial^{k}u_{1}}{\partial x^{k}}(\ell,y,t) &= \sum_{k=0}^{\infty} \frac{\varepsilon^{k}w(y)^{k}}{k!} \frac{\partial^{k}u_{2}}{\partial x^{k}}(\ell,y,t),
\end{align*}
and inserting the expansions (\ref{eq:ansatz1}) and (\ref{eq:ansatz2}) yields:
\begin{align}
\label{eq:int1_expansion} 
\sum_{k=0}^{\infty}\sum_{i=0}^{\infty}\frac{\varepsilon^{i+k}w(y)^{k}}{k!}\frac{\partial^{k}u_{1}^{(i)}}{\partial x^{k}}(\ell,y,t) = \sum_{k=0}^{\infty}\sum_{i=0}^{\infty}\frac{\varepsilon^{i+k}w(y)^{k}}{k!}\frac{\partial^{k}u_{2}^{(i)}}{\partial x^{k}}(\ell,y,t).
\end{align}
For all $n\in\mathbb{N}$, the $\mathcal{O}(\varepsilon^{n})$ term on both sides of (\ref{eq:int1_expansion}) is identified when $i + k = n$ or equivalently, $i = n-k$ for $k = 0,\hdots,n$. Matching the $\mathcal{O}(\varepsilon^{n})$ terms on both sides of (\ref{eq:int1_expansion}) therefore yields:
\begin{align*}
\sum_{k=0}^{n}\frac{w(y)^{k}}{k!}\frac{\partial^{k}u_{1}^{(n-k)}}{\partial x^{k}}(\ell,y,t) = \sum_{k=0}^{n}\frac{w(y)^{k}}{k!}\frac{\partial^{k}u_{2}^{(n-k)}}{\partial x^{k}}(\ell,y,t).
\end{align*}
Hence, we have derived the following interface condition satisfied by $u_{1}^{(n)}$ and $u_{2}^{(n)}$ for all $n\in\mathbb{N}$:
\begin{gather*}
u_{1}^{(n)}(\ell,y,t) = u_{2}^{(n)}(\ell,y,t) + f^{(n)}(y,t),
\end{gather*}
where $f^{(n)}(y,t)$ depends on (derivatives of) $u_{1}^{(0)},\hdots,u_{1}^{(n-1)}$ and $u_{2}^{(0)},\hdots,u_{2}^{(n-1)}$:
\begin{gather}
\label{eq:fn}
f^{(n)}(y,t) = \sum_{k=1}^{n}\frac{w(y)^{k}}{k!}\left[\frac{\partial^{k}u_{2}^{(n-k)}}{\partial x^{k}}(\ell,y,t)-\frac{\partial^{k}u_{1}^{(n-k)}}{\partial x^{k}}(\ell,y,t)\right]\!.
\end{gather}

Next consider the interface condition (\ref{eq:int2}). Expanding the first derivatives of $u_{1}$ and $u_{2}$ in Taylor series centered at $x = \ell$ and then inserting the expansions (\ref{eq:ansatz1}) and (\ref{eq:ansatz2}) yields:
\begin{align*}
\frac{\partial u_{1}}{\partial x}(\ell+\varepsilon w(y),y,t) &= \sum_{k=0}^{\infty}\sum_{i=0}^{\infty}\frac{\varepsilon^{i+k}w(y)^{k}}{k!}\frac{\partial^{k+1}u_{1}^{(i)}}{\partial x^{k+1}}(\ell,y,t),\\
\frac{\partial u_{1}}{\partial y}(\ell+\varepsilon w(y),y,t) &= \sum_{k=0}^{\infty}\sum_{i=0}^{\infty}\frac{\varepsilon^{i+k}w(y)^{k}}{k!}\frac{\partial^{k+1}u_{1}^{(i)}}{\partial x^{k}\partial y}(\ell,y,t),\\
\frac{\partial u_{2}}{\partial x}(\ell+\varepsilon w(y),y,t) &= \sum_{k=0}^{\infty}\sum_{i=0}^{\infty}\frac{\varepsilon^{i+k}w(y)^{k}}{k!}\frac{\partial^{k+1}u_{2}^{(i)}}{\partial x^{k+1}}(\ell,y,t),\\
\frac{\partial u_{2}}{\partial y}(\ell+\varepsilon w(y),y,t) &= \sum_{k=0}^{\infty}\sum_{i=0}^{\infty}\frac{\varepsilon^{i+k}w(y)^{k}}{k!}\frac{\partial^{k+1}u_{2}^{(i)}}{\partial x^{k}\partial y}(\ell,y,t).
\end{align*}
The perturbed interface can be described by the vector-valued function $\mathbf{v}(y) = (\ell+\varepsilon w(y),y)$ with tangent vector $\mathbf{v}'(y) = (\varepsilon w'(y),1)$, which allows a normal vector to be identified, $\mathbf{n}(y) = (1,-\varepsilon w'(y))$. Note that $\mathbf{n}(y)$ is not required to have unit length as the magnitude of $\mathbf{n}(y)$ does not affect the interface condition (\ref{eq:int2}). Similarly we have chosen $\mathbf{n}(y)$ to point outwards from the first region but in the working that follows one could just as easily use the normal vector pointing inwards, as a negative sign also does not affect equation (\ref{eq:int2}). Combining the form of $\mathbf{n}(y)$ with the above expressions for the first derivatives yields the following expansion of the interface condition (\ref{eq:int2}):
\begin{multline}
\label{eq:int2_expansion}
D_{1}\left[\sum_{k=0}^{\infty}\sum_{i=0}^{\infty}\frac{\varepsilon^{i+k}w(y)^{k}}{k!}\frac{\partial^{k+1}u_{1}^{(i)}}{\partial x^{k+1}}(\ell,y,t) - w'(y)\sum_{k=0}^{\infty}\sum_{i=0}^{\infty}\frac{\varepsilon^{i+k+1}w(y)^{k}}{k!}\frac{\partial^{k+1}u_{1}^{(i)}}{\partial x^{k}\partial y}(\ell,y,t)\right]\!\\ = D_{2}\left[\sum_{k=0}^{\infty}\sum_{i=0}^{\infty}\frac{\varepsilon^{i+k}w(y)^{k}}{k!}\frac{\partial^{k+1}u_{2}^{(i)}}{\partial x^{k+1}}(\ell,y,t) - w'(y)\sum_{k=0}^{\infty}\sum_{i=0}^{\infty}\frac{\varepsilon^{i+k+1}w(y)^{k}}{k!}\frac{\partial^{k+1}u_{2}^{(i)}}{\partial x^{k}\partial y}(\ell,y,t)\right]\!.
\end{multline}
Here, for all $n\in\mathbb{N}$, the $\mathcal{O}(\varepsilon^{n})$ terms on both sides of (\ref{eq:int2_expansion}) are identified when $i + k = n$ in the first double sum (or equivalently, $i = n-k$ for $k = 0,\hdots,n$) and when $i + k + 1 = n$ in the second double sum (or equivalently, $i = n-k-1$ for $k = 0,\hdots,n-1$). Thus matching the $\mathcal{O}(\varepsilon^{n})$ terms on both sides of (\ref{eq:int2_expansion}) yields:
\begin{multline*}
D_{1}\left[\sum_{k=0}^{n}\frac{w(y)^{k}}{k!}\frac{\partial^{k+1}u_{1}^{(n-k)}}{\partial x^{k+1}}(\ell,y,t) - w'(y)\sum_{k=0}^{n-1}\frac{w(y)^{k}}{k!}\frac{\partial^{k+1}u_{1}^{(n-k-1)}}{\partial x^{k}\partial y}(\ell,y,t)\right]\\ = D_{2}\left[\sum_{k=0}^{n}\frac{w(y)^{k}}{k!}\frac{\partial^{k+1}u_{2}^{(n-k)}}{\partial x^{k+1}}(\ell,y,t) - w'(y)\sum_{k=0}^{n-1}\frac{w(y)^{k}}{k!}\frac{\partial^{k+1}u_{2}^{(n-k-1)}}{\partial x^{k}\partial y}(\ell,y,t)\right]\!.
\end{multline*}
Hence, we have derived the following interface condition satisfied by $u_{1}^{(n)}$ and $u_{2}^{(n)}$ for all $n\in\mathbb{N}$:
\begin{gather*}
D_{1}\frac{\partial u_{1}^{(n)}}{\partial x}(\ell,y,t) = D_{2}\frac{\partial u_{2}^{(n)}}{\partial x}(\ell,y,t) + g^{(n)}(y,t),
\end{gather*}
where, like $f^{(n)}(y,t)$, the function $g^{(n)}(y,t)$ depends on derivatives of $u_{1}^{(0)},\hdots,u_{1}^{(n-1)}$ and $u_{2}^{(0)},\hdots,u_{2}^{(n-1)}$:
\begin{multline}
\label{eq:gn}
g^{(n)}(y,t) = \sum_{k=1}^{n}\frac{w(y)^{k}}{k!}\left[D_{2}\frac{\partial^{k+1}u_{2}^{(n-k)}}{\partial x^{k+1}}(\ell,y,t)-D_{1}\frac{\partial^{k+1}u_{1}^{(n-k)}}{\partial x^{k+1}}(\ell,y,t)\right]\\ + w'(y)\sum_{k=0}^{n-1}\frac{w(y)^{k}}{k!}\left[D_{1}\frac{\partial^{k+1}u_{1}^{(n-k-1)}}{\partial x^{k}\partial y}(\ell,y,t) - D_{2}\frac{\partial^{k+1}u_{1}^{(n-k-1)}}{\partial x^{k}\partial y}(\ell,y,t)\right]\!.
\end{multline}

In summary, $u_{1}^{(n)}$ and $u_{2}^{(n)}$ satisfy the following partial differential equations and interface conditions:
\begin{gather}
\label{eq:ordern_pde1}
\frac{\partial u_{1}^{(n)}}{\partial t} = D_{1}\Delta u_{1}^{(n)},\quad 0 < x < \ell,\enspace 0 < y < H,\enspace t > 0,\\
\label{eq:ordern_pde2}
\frac{\partial u_{2}^{(n)}}{\partial t} = D_{2}\Delta u_{2}^{(n)},\quad \ell < x < L,\enspace 0 < y < H,\enspace t > 0,\\
\label{eq:ordern_int1}
u_{1}^{(n)}(\ell,y,t) = u_{2}^{(n)}(\ell,y,t) + f^{(n)}(y,t),\quad 0 < y < H,\enspace t > 0,\\
\label{eq:ordern_int2}
D_{1}\frac{\partial u_{1}^{(n)}}{\partial x}(\ell,y,t) = D_{2}\frac{\partial u_{2}^{(n)}}{\partial x}(\ell,y,t) + g^{(n)}(y,t),\quad 0 < y < H,\enspace t > 0,
\end{gather}
for all $n\in\mathbb{N}$. Here, the \revision{interface conditions} (\ref{eq:ordern_int1}) and (\ref{eq:ordern_int2}) can be interpreted as continuity of solution and flux (in the $x$ direction) at the unperturbed interface ($x = \ell$) with correction terms $f^{(n)}(y,t)$ and $g^{(n)}(y,t)$ due to the perturbed interface. The above analysis has converted the problem on the perturbed domain into a sequence of problems on the unperturbed domain. The tradeoff being that the interface conditions (\ref{eq:ordern_int1}) and (\ref{eq:ordern_int2}) now involve non-homogeneities. 

\section{Test Case}
\label{sec:test_case}
So far we have neglected specifying boundary conditions at the external boundaries ($x = 0, L$ and $y = 0,H$). To demonstrate the asymptotic expansion solution for equations (\ref{eq:pde1})--(\ref{eq:int2}), we now consider the specific case of the following initial and boundary conditions, which are motivated by those featuring in mathematical models of heat transfer in skin \cite{mcinerney_2019,simpson_2017} with heat applied at $x = 0$, (possible) heat loss at $x = L$ and zero heat flux at $y = 0,H$:
\begin{gather}
\label{eq:ic1}
u_{1}(x,y,0) = 0,\quad 0 < x < \ell,\enspace 0 < y < H,\\
\label{eq:ic2}
u_{2}(x,y,0) = 0,\quad \ell < x < L,\enspace 0 < y < H,\\
\label{eq:bc1}
u_{1}(0,y,t) = c_{0}(t),\quad \frac{\partial u_{2}}{\partial x}(L,y,t) = \revision{q_{L}}(t),\quad 0 < y < H,\enspace t > 0,\\
\label{eq:bc2a}
\frac{\partial u_{1}}{\partial y}(x,0,t) = 0,\quad\frac{\partial u_{1}}{\partial y}(x,H,t) = 0,\quad 0 < x < \ell + \varepsilon w(y),\enspace t > 0,\\
\label{eq:bc2b}
\frac{\partial u_{2}}{\partial y}(x,0,t) = 0,\quad\frac{\partial u_{2}}{\partial y}(x,H,t) = 0,\quad \ell + \varepsilon w(y) < x < L,\enspace t > 0.
\end{gather}
Substituting the expansions (\ref{eq:ansatz1}) and (\ref{eq:ansatz2}) into the above initial and boundary conditions (\ref{eq:ic1})--(\ref{eq:bc2b}) and matching powers of the $\mathcal{O}(\varepsilon^{n})$ terms yields the appropriate initial and boundary conditions for $u_{1}^{(n)}$ and $u_{2}^{(n)}$. Combining these conditions with equations (\ref{eq:ordern_pde1})--(\ref{eq:ordern_int2}) yields the following initial-boundary value problem for all $n\in\mathbb{N}$:
\begin{gather}
\label{eq:ordern2_pde1}
\frac{\partial u_{1}^{(n)}}{\partial t} = D_{1}\Delta u_{1}^{(n)},\quad 0 < x < \ell,\enspace 0 < y < H,\enspace t > 0,\\
\label{eq:ordern2_pde2}
\frac{\partial u_{2}^{(n)}}{\partial t} = D_{2}\Delta u_{2}^{(n)},\quad \ell < x < L,\enspace 0 < y < H,\enspace t > 0,\\
\label{eq:ordern2_ic1}
u_{1}^{(n)}(x,y,0) = 0,\quad 0 < x < \ell,\enspace 0 < y < H,\\
\label{eq:ordern2_ic2}
u_{2}^{(n)}(x,y,0) = 0,\quad \ell < x < L,\enspace 0 < y < H,\\
\label{eq:ordern2_bc1}
u_{1}^{(n)}(0,y,t) = \begin{cases} c_{0}(t), & \text{if $n = 0$,}\\ 0, & \text{if $n\in\mathbb{N}^{+}$,}\end{cases}\quad \frac{\partial u_{2}^{(n)}}{\partial x}(L,y,t) = \begin{cases} \revision{q_{L}}(t), & \text{if $n = 0$,}\\ 0, & \text{if $n\in\mathbb{N}^{+}$,}\end{cases}\quad 0 < y < H,\enspace t > 0,\\
\label{eq:ordern2_bc2}
\frac{\partial u_{1}^{(n)}}{\partial y}(x,0,t) = 0,\quad\frac{\partial u_{1}^{(n)}}{\partial y}(x,H,t) = 0,\quad 0 < x < \ell,\enspace t > 0,\\
\label{eq:ordern2_bc3}
\frac{\partial u_{2}^{(n)}}{\partial y}(x,0,t) = 0,\quad\frac{\partial u_{2}^{(n)}}{\partial y}(x,H,t) = 0,\quad \ell < x < L,\enspace t > 0,\\
\label{eq:ordern2_int1}
u_{1}^{(n)}(\ell,y,t) = u_{2}^{(n)}(\ell,y,t) + f^{(n)}(y,t),\quad 0 < y < H,\enspace t > 0,\\
\label{eq:ordern2_int2}
D_{1}\frac{\partial u_{1}^{(n)}}{\partial x}(\ell,y,t) = D_{2}\frac{\partial u_{2}^{(n)}}{\partial x}(\ell,y,t) + g^{(n)}(y,t),\quad 0 < y < H,\enspace t > 0,
\end{gather}
where $\mathbb{N}^{+} := \{1,2,\hdots\}$. We remark that the initial-boundary value problem for $u_{1}^{(n)}$ and $u_{2}^{(n)}$ depends on $u_{1}^{(0)},u_{2}^{(0)},\hdots,u_{1}^{(n-1)},u_{2}^{(n-1)}$ (through $f^{(n)}(y,t)$ (\ref{eq:fn}) and $g^{(n)}(y,t)$ (\ref{eq:gn})) and therefore each problem must be solved for $n = 0,1,2,\hdots$ sequentially.  

In this work, we solve the initial-boundary value problem (\ref{eq:ordern2_pde1})--(\ref{eq:ordern2_int2}) using the Laplace transform. Let $U_{1}^{(n)}(x,y,s) = \mathcal{L}\{u_{1}^{(n)}(x,y,t)\}$ and $U_{2}^{(n)}(x,y,s) = \mathcal{L}\{u_{2}^{(n)}(x,y,t)\}$ be the Laplace transformations of $u_{1}^{(n)}(x,y,t)$ and $u_{2}^{(n)}(x,y,t)$ with respect to $t$, respectively, where $s$ is the transformation variable. Taking the Laplace transform of (\ref{eq:ordern2_pde1})--(\ref{eq:ordern2_int2}) yields the boundary value problem:
\begin{gather}
\label{eq:ordern_pde1_lt}
sU_{1}^{(n)} = D_{1}\Delta U_{1}^{(n)},\quad 0 < x < \ell,\enspace 0 < y < H,\\
\label{eq:ordern_pde2_lt}
sU_{2}^{(n)} = D_{2}\Delta U_{2}^{(n)},\quad \ell < x < L,\enspace 0 < y < H,\\
\label{eq:ordern_bc1_lt}
U_{1}^{(n)}(0,y,s) = \begin{cases} C_{0}(s), & \text{if $n = 0$,}\\ 0, & \text{if $n\in\mathbb{N}^{+}$,}\end{cases}\quad \frac{\partial U_{2}^{(n)}}{\partial x}(L,y,s) = \begin{cases} \revision{Q_{L}}(s), & \text{if $n = 0$,}\\ 0, & \text{if $n\in\mathbb{N}^{+}$,}\end{cases}\quad 0 < y < H,\\
\label{eq:ordern_bc2_lt}
\frac{\partial U_{1}^{(n)}}{\partial y}(x,0,s) = 0,\quad\frac{\partial U_{1}^{(n)}}{\partial y}(x,H,s) = 0,\quad 0 < x < \ell,\\
\label{eq:ordern_bc3_lt}
\frac{\partial U_{2}^{(n)}}{\partial y}(x,0,s) = 0,\quad\frac{\partial U_{2}^{(n)}}{\partial y}(x,H,s) = 0,\quad \ell < x < L,\\
\label{eq:ordern_int1_lt}
U_{1}^{(n)}(\ell,y,s) = U_{2}^{(n)}(\ell,y,s) + F^{(n)}(y,s),\quad 0 < y < H,\\
\label{eq:ordern_int2_lt}
D_{1}\frac{\partial U_{1}^{(n)}}{\partial x}(\ell,y,s) = D_{2}\frac{\partial U_{2}^{(n)}}{\partial x}(\ell,y,s) + G^{(n)}(y,s),\quad 0 < y < H,
\end{gather}
for all $n\in\mathbb{N}$, where $C_{0}(s) = \mathcal{L}\{c_{0}(t)\}$, $\revision{Q_{L}}(s) = \mathcal{L}\{\revision{q_{L}}(t)\}$ and $F^{(n)}(y,s)$ and $G^{(n)}(y,s)$ are the Laplace transformations of $f^{(n)}(y,t)$ (\ref{eq:fn}) and $g^{(n)}(y,t)$ (\ref{eq:gn}):
\begin{gather}
\label{eq:Fn}
F^{(n)}(y,s) = \sum_{k=1}^{n}\frac{w(y)^{k}}{k!}\left[\frac{\partial^{k}U_{2}^{(n-k)}}{\partial x^{k}}(\ell,y,s)-\frac{\partial^{k}U_{1}^{(n-k)}}{\partial x^{k}}(\ell,y,s)\right]\!,\\
\nonumber
G^{(n)}(y,s) = \sum_{k=1}^{n}\frac{w(y)^{k}}{k!}\left[D_{2}\frac{\partial^{k+1}U_{2}^{(n-k)}}{\partial x^{k+1}}(\ell,y,s)-D_{1}\frac{\partial^{k+1}U_{1}^{(n-k)}}{\partial x^{k+1}}(\ell,y,s)\right]\\ 
\label{eq:Gn}
\hspace*{7em} + w'(y)\sum_{k=0}^{n-1}\frac{w(y)^{k}}{k!}\left[D_{1}\frac{\partial^{k+1}U_{1}^{(n-k-1)}}{\partial x^{k}\partial y}(\ell,y,s) - D_{2}\frac{\partial^{k+1}U_{2}^{(n-k-1)}}{\partial x^{k}\partial y}(\ell,y,s)\right]\!.
\end{gather}
In the following subsections we obtain expressions for $U_{1}^{(n)}$ and $U_{2}^{(n)}$, considering the cases of $n = 0$ and $n\in\mathbb{N}^{+}$ separately.

\subsection{Leading order term}
\label{sec:leading_order}
Consider the initial-boundary value problem (\ref{eq:ordern_pde1_lt})--(\ref{eq:ordern_int2_lt}) for $n = 0$. Due to the boundary conditions (\ref{eq:ordern2_bc2}) and (\ref{eq:ordern2_bc3}) and the fact that $f^{(0)}(y,t) = g^{(0)}(y,t) = 0$, both $u_{1}^{(0)}$ and $u_{2}^{(0)}$ are independent of $y$ and we write $u_{1}^{(0)}(x,y,t) \equiv u_{1}^{(0)}(x,t)$ and $u_{2}^{(0)}(x,y,t) \equiv u_{2}^{(0)}(x,t)$ and hence $U_{1}^{(n)}(x,y,s) \equiv U_{1}^{(n)}(x,s)$ and $U_{2}^{(n)}(x,y,s) \equiv U_{2}^{(n)}(x,s)$. It follows that the initial-boundary value problem (\ref{eq:ordern_pde1_lt})--(\ref{eq:ordern_int2_lt}) for $n = 0$ reduces to the following two-layer one-dimensional problem:
\begin{gather}
\label{eq:order0_pde1}
sU_{1}^{(0)} = D_{1}\frac{\partial^{2} U_{1}^{(0)}}{\partial x^{2}},\quad 0 < x < \ell,\\
\label{eq:order0_pde2}
sU_{2}^{(0)} = D_{2}\frac{\partial^{2} U_{2}^{(0)}}{\partial x^{2}},\quad \ell < x < L,\\
\label{eq:order0_bc1}
U_{1}^{(0)}(0,s) = C_{0}(s),\quad \frac{\partial U_{2}^{(0)}}{\partial x}(L,s) = \revision{Q_{L}}(s),\\
\label{eq:order0_int1}
U_{1}^{(0)}(\ell,s) = U_{2}^{(0)}(\ell,s),\\
\label{eq:order0_int2}
D_{1}\frac{\partial U_{1}^{(0)}}{\partial x}(\ell,s) = D_{2}\frac{\partial U_{2}^{(0)}}{\partial x}(\ell,s).
\end{gather}
To solve (\ref{eq:order0_pde1})--(\ref{eq:order0_int2}), we reformulate the problem by setting
\begin{align*}
V^{(0)}(s) := D_{2}\frac{\partial U_{2}^{(0)}}{\partial x}(\ell,s),
\end{align*}
which gives standard boundary value problems on each layer:

\bigskip
\noindent\textit{Layer 1:}
\begin{gather}
\label{eq:order0_layer1_pde1}
sU_{1}^{(0)} = D_{1}\frac{\partial^{2} U_{1}^{(0)}}{\partial x^{2}},\quad 0 < x < \ell,\\
\label{eq:order0_layer1_bc1}
U_{1}^{(0)}(0,s) = C_{0}(s),\quad D_{1}\frac{\partial U_{1}^{(0)}}{\partial x}(\ell,s) = V^{(0)}(s).
\end{gather}
\textit{Layer 2:}
\begin{gather}
\label{eq:order0_layer2_pde1}
sU_{2}^{(0)} = D_{2}\frac{\partial^{2} U_{2}^{(0)}}{\partial x^{2}},\quad \ell < x < L,\\
\label{eq:order0_layer2_bc1}
D_{2}\frac{\partial U_{2}^{(0)}}{\partial x}(\ell,s) = V^{(0)}(s),\quad \frac{\partial U_{2}^{(0)}}{\partial x}(L,s) = \revision{Q_{L}}(s).
\end{gather}
The boundary value problems (\ref{eq:order0_layer1_pde1})--(\ref{eq:order0_layer1_bc1}) and (\ref{eq:order0_layer2_pde1})--(\ref{eq:order0_layer2_bc1}) involve second-order constant coefficient differential equations and thus can be solved using standard techniques to give:
\begin{align}
\label{eq:U10}
U_{1}^{(0)}(x,s) &= \left[\gamma_{11}(s) + \gamma_{12}(s)V^{(0)}(s)\right]\exp(\mu_{1}(s)x) + \left[\gamma_{13}(s) + \gamma_{14}(s)V^{(0)}(s)\right]\exp(-\mu_{1}(s)x),\\
\label{eq:U20}
U_{2}^{(0)}(x,s) &= \left[\gamma_{21}(s) + \gamma_{22}(s)V^{(0)}(s)\right]\exp(\mu_{2}(s)x) + \left[\gamma_{23}(s) + \gamma_{24}(s)V^{(0)}(s)\right]\exp(-\mu_{2}(s)x),
\end{align}
where all variables (except $V^{(0)}(s)$) are defined in the Appendix and $V^{(0)}(s)$ is identified by enforcing the interface condition (\ref{eq:order0_int1}), which was absent in the reformulated problems (\ref{eq:order0_layer1_pde1})--(\ref{eq:order0_layer1_bc1}) and (\ref{eq:order0_layer2_pde1})--(\ref{eq:order0_layer2_bc1}), yielding:
\begin{align*}
V^{(0)}(s) = \frac{\gamma_{11}(s)\exp(\mu_{1}(s)\ell) + \gamma_{13}(s)\exp(-\mu_{1}(s)\ell) - \gamma_{21}(s)\exp(\mu_{2}(s)\ell) - \gamma_{23}(s)\exp(-\mu_{2}(s)\ell)}{\gamma_{22}(s)\exp(\mu_{2}(s)\ell) + \gamma_{24}(s)\exp(-\mu_{2}(s)\ell) - \gamma_{12}(s)\exp(\mu_{1}(s)\ell) - \gamma_{14}(s)\exp(-\mu_{1}(s)\ell)}.
\end{align*}
In summary, with $V^{(0)}(s)$ now known, both $U_{1}^{(0)}(x,s)$ (\ref{eq:U10}) and $U_{2}^{(0)}(x,s)$ (\ref{eq:U20}) are fully identified. 

\subsection{Higher order terms}
Consider the boundary value problem (\ref{eq:ordern_pde1_lt})--(\ref{eq:ordern_int2_lt}) for $n\in\mathbb{N}^{+}$. To solve this problem we again reformulate the problem by setting
\begin{align*}
V^{(n)}(y,s) = D_{2}\frac{\partial U_{2}^{(n)}}{\partial x}(\ell,y,s),
\end{align*}
which gives standard initial-boundary value problems on each layer for all $n\in\mathbb{N}^{+}$:

\bigskip
\noindent\textit{Layer 1:}
\begin{gather}
\label{eq:ordern_layer1_pde1}
sU_{1}^{(n)} = D_{1}\Delta U_{1}^{(n)},\quad 0 < x < \ell,\enspace 0 < y < H,\\
\label{eq:ordern_layer1_bc1}
U_{1}^{(n)}(0,y,s) = 0,\quad D_{1}\frac{\partial U_{1}^{(n)}}{\partial x}(\ell,y,s) = V^{(n)}(y,s) + G^{(n)}(y,s),\quad 0 < y < H,\\
\label{eq:ordern_layer1_bc2}
\frac{\partial U_{1}^{(n)}}{\partial y}(x,0,s) = 0,\quad \frac{\partial U_{1}^{(n)}}{\partial y}(x,H,s) = 0,\quad 0 < x < \ell.
\end{gather}
\textit{Layer 2:}
\begin{gather}
\label{eq:ordern_layer2_pde1}
sU_{2}^{(n)} = D_{2}\Delta U_{2}^{(n)},\quad \ell < x < L,\enspace 0 < y < H,\\
\label{eq:ordern_layer2_bc1}
D_{2}\frac{\partial U_{2}^{(n)}}{\partial x}(\ell,y,s) = V^{(n)}(y,s),\quad \frac{\partial U_{2}^{(n)}}{\partial x}(L,y,s) = 0,\quad 0 < y < H,\\
\label{eq:ordern_layer2_bc2}
\frac{\partial U_{2}^{(n)}}{\partial y}(x,0,s) = 0,\quad \frac{\partial U_{2}^{(n)}}{\partial y}(x,H,s) = 0,\quad \ell < x < L.
\end{gather}
Both boundary value problems (\ref{eq:ordern_layer1_pde1})--(\ref{eq:ordern_layer1_bc2}) and (\ref{eq:ordern_layer2_pde1})--(\ref{eq:ordern_layer2_bc2}) can be solved using the  standard techniques of separation of variables and eigenfunction expansion to give:
\begin{align}
\label{eq:U1n}
U_{1}^{(n)}(x,y,s) &= \sum_{m=0}^{\infty} \alpha_{m}^{(n)}(s)\sinh(\mu_{1,m}(s)x)\cos(\lambda_{m}y),\\
\label{eq:U2n}
U_{2}^{(n)}(x,y,s) &= \sum_{m=0}^{\infty} \beta_{m}^{(n)}(s)\cosh(\mu_{2,m}(s)[x-L])\cos(\lambda_{m}y),
\end{align}
where
\begin{gather*}
\alpha_{m}^{(n)}(s) = \widetilde{\alpha}_{m}^{(n)}(s)\widetilde{V}_{m}^{(n)}(s),\quad\beta_{m}^{(n)}(s) = \widetilde{\beta}_{m}^{(n)}(s)[\widetilde{V}_{m}^{(n)}(s) - \widetilde{G}_{m}^{(n)}(s)],\\\widetilde{V}_{m}^{(n)}(s) = \int_{0}^{H}V^{(n)}(y,s)\cos(\lambda_{m}y)\,\text{d}y,\quad \widetilde{G}_{m}^{(n)}(s) = \int_{0}^{H}G^{(n)}(y,s)\cos(\lambda_{m}y)\,\text{d}y,
\end{gather*}
and all other remaining variables are defined in the Appendix. Here, we see that $V^{(n)}(y,s)$ is not explicitly required, only $\widetilde{V}_{m}^{(n)}(s)$ for $m\in\mathbb{N}^{+}$, the latter of which is identified by enforcing that $U_{1}^{(n)}$ (\ref{eq:U1n}) and $U_{2}^{(n)}$ (\ref{eq:U2n}) satisfy the interface condition (\ref{eq:ordern_int1_lt}) and using orthogonality of the eigenfunctions $\cos(\lambda_{m}y)$ to give:
\begin{align*}
\widetilde{V}_{m}^{(n)}(s) = \begin{cases} \dfrac{\frac{1}{H}\widetilde{F}_{m}^{(n)}(s) - \widetilde{\beta}_{m}^{(n)}(s)\widetilde{G}_{m}^{(n)}(s)\cosh(\mu_{2,m}(s)[\ell-L])}{\widetilde{\alpha}_{m}^{(n)}(s)\sinh(\mu_{1,m}(s)\ell) - \widetilde{\beta}_{m}^{(n)}(s)\cosh(\mu_{2,n}[\ell-L])}, & \text{if $m = 0$},\\ \dfrac{\frac{2}{H}\widetilde{F}_{m}^{(n)}(s) - \widetilde{\beta}_{m}^{(n)}(s)\widetilde{G}_{m}^{(n)}(s)\cosh(\mu_{2,m}(s)[\ell-L])}{\widetilde{\alpha}_{m}^{(n)}(s)\sinh(\mu_{1,m}(s)\ell) - \widetilde{\beta}_{m}^{(n)}(s)\cosh(\mu_{2,n}[\ell-L])}, & \text{if $m\in\mathbb{N}^{+}$},\end{cases}
\end{align*}
where
\begin{gather*}
\widetilde{F}_{m}^{(n)}(s) = \int_{0}^{H}F^{(n)}(y,s)\cos(\lambda_{m}y)\,\text{d}y.
\end{gather*}
In summary, with $V_{m}^{(n)}(s)$ now known for all $m\in\mathbb{N}$, both $U_{1}^{(n)}$ (\ref{eq:U1n}) and $U_{2}^{(n)}$ (\ref{eq:U2n}) are fully identified. The last thing to address is computation of the integral expressions for $\widetilde{F}_{m}^{(n)}(s)$ and $\widetilde{G}_{m}^{(n)}(s)$ which involve the functions $F^{(n)}(y,s)$ (\ref{eq:Fn}) and $G^{(n)}(y,s)$ (\ref{eq:Gn}). The derivatives appearing in $F^{(n)}(y,s)$ (\ref{eq:Fn}) and $G^{(n)}(y,s)$ (\ref{eq:Gn}) are calculated exactly and given in the Appendix. While the integral expressions for $\widetilde{F}_{m}^{(n)}(s)$ and $\widetilde{G}_{m}^{(n)}(s)$ can be calculated exactly for specific choices of $w(y)$, we use MATLAB's \verb"integral" function to keep our code as general as possible.

\subsection{Inverse Laplace transform} 
\label{sec:inverse_laplace_transform}
With $U_{1}^{(0)}$, $U_{2}^{(0)}$, $U_{1}^{(n)}$ for $n\in\mathbb{N}^{+}$, and $U_{2}^{(n)}$ for $n\in\mathbb{N}^{+}$ now known, the final step is to transform the solution from the Laplace domain back to the time domain. \revision{This yields the following perturbation solution of (\ref{eq:pde1})--(\ref{eq:int2}) and (\ref{eq:ic1})--(\ref{eq:bc2b}):}
\begin{align}
\label{eq:u1}
u_{1}(x,y,t) = \mathcal{L}^{-1}\left\{U_{1}^{(0)}(x,s)\right\} + \sum_{n=1}^{N-1} \varepsilon^{n}\mathcal{L}^{-1}\left\{U_{1}^{(n)}(x,y,s)\right\}\!,\\
\label{eq:u2}
u_{2}(x,y,t) = \mathcal{L}^{-1}\left\{U_{2}^{(0)}(x,s)\right\} + \sum_{n=1}^{N-1} \varepsilon^{n}\mathcal{L}^{-1}\left\{U_{2}^{(n)}(x,y,s)\right\}\!,
\end{align}
\revision{when truncating the expansions at a finite number of terms, $N$. Due to the complicated form of $U_{1}^{(n)}$ (\ref{eq:U1n}) and $U_{2}^{(n)}$ (\ref{eq:U2n}) and the fact that they depend recursively on the previous terms $U_{1}^{(0)},\hdots,U_{1}^{(n-1)}$ and $U_{2}^{(0)},\hdots,U_{2}^{(n-1)}$ (through $F^{(n)}(y,t)$ (\ref{eq:Fn}) and $G^{(n)}(y,t)$ (\ref{eq:Gn})), analytically inverting the Laplace transforms in (\ref{eq:u1}) and (\ref{eq:u2}) is very challenging.} To address this, we use a numerical inverse Laplace transform approximation by \citet{trefethen_2006}, which has frequently been used for other similar heterogeneous transport problems (e.g. \cite{carr_2016,ilic_2010,carr_2021}). 

\revision{During our numerical investigations, we found that both $\mathcal{L}^{-1}\{U_{1}^{(n)}(x,y,s)\}$ and $\mathcal{L}^{-1}\{U_{2}^{(n)}(x,y,s)\}$ tend to increase in magnitude as $n$ increases. However, provided $\varepsilon$ is sufficiently small this is balanced out by the increasing powers of $\varepsilon$ in the expansions (\ref{eq:u1}) and (\ref{eq:u2}). It is also important to note here that the expansions (\ref{eq:u1}) and (\ref{eq:u2}) converge in the limit as $\varepsilon\rightarrow 0$ but not necessarily in the limit as $N\rightarrow\infty$. This means that for a fixed value of $\varepsilon$, increasing $N$ does not necessarily improve the accuracy of the perturbation solution. As we will see in the next section, the usefulness comes from the fact that expansions (\ref{eq:u1}) and (\ref{eq:u2}) produce accurate approximations to the solution for small values of $N$.}

\revision{\section{Results}}
\label{sec:results}
We now compare our semi-analytical perturbation solution (developed in section \ref{sec:test_case}) to a numerical solution obtained using a standard finite volume spatial discretisation (described in detail in previous work \cite{pontrelli_2020,simpson_2021}). In all results, we compare both solutions at $t = 0.2$ for $D_{1} = 1$, $D_{2} = 0.01$, $L = H = 1$, $\ell = 0.5$, \revision{$c_{0}(t) = 1$, $q_{L}(t) = 0$} and four different combinations of $\varepsilon$ and $w(y)$. In all cases, the perturbation solution is computed using the first $30$ terms in the eigenfunction expansions (\ref{eq:U1n})--(\ref{eq:U2n}) and (\ref{eq:diffU1n})--(\ref{eq:diffU2n}). Both the perturbation and numerical solutions are evaluated/computed at the vertices of an unstructured triangular mesh conforming to the perturbed domain (Figure \ref{fig:skin}(b)). Here vertices are positioned along the interface at $x = \ell + \varepsilon w(y)$ and each triangular element is located entirely within either region $1$ or $2$. All meshes are generated using GMSH \cite{gmsh_2009} with refinement controlled by setting a mesh element size of 0.01, which yields a mesh consisting of approximately 12,000 nodes and 24,000 elements for all geometries tested. Complete details of our code implementation and experiments can be found in our MATLAB code available at \href{https://github.com/elliotcarr/Carr2022a}{https://github.com/elliotcarr/Carr2022a}. 

\begin{figure}[!t]
\centering
\includegraphics[width=0.97\textwidth]{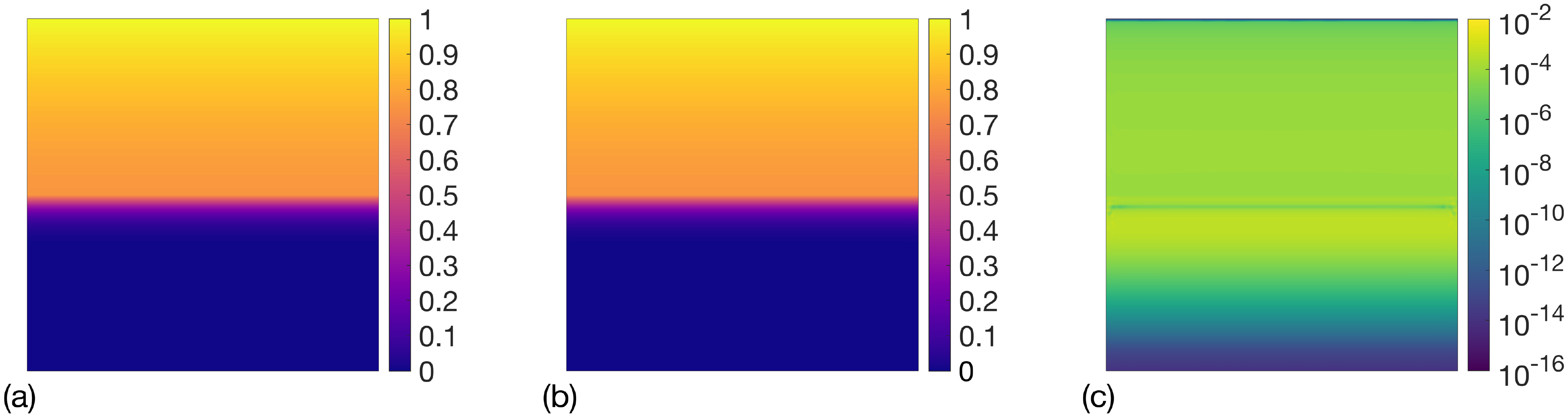}
\includegraphics[width=0.97\textwidth]{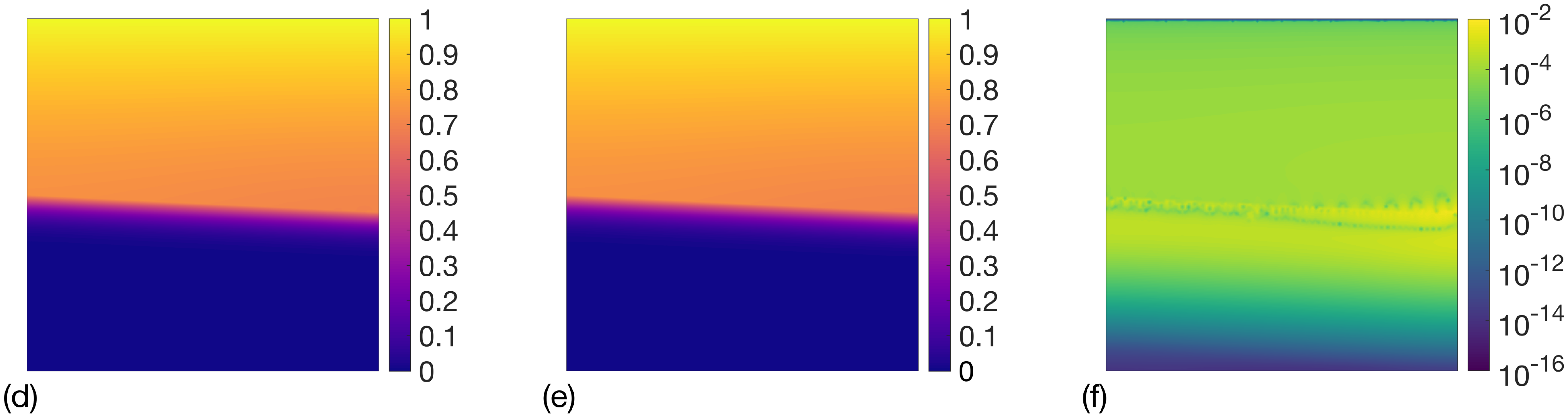}
\includegraphics[width=0.97\textwidth]{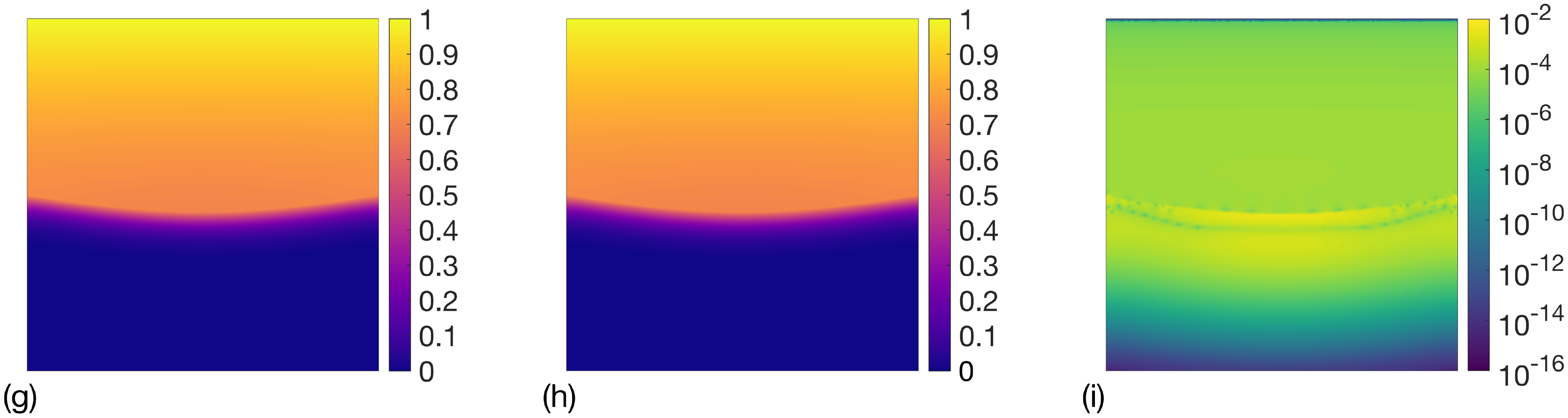}
\includegraphics[width=0.97\textwidth]{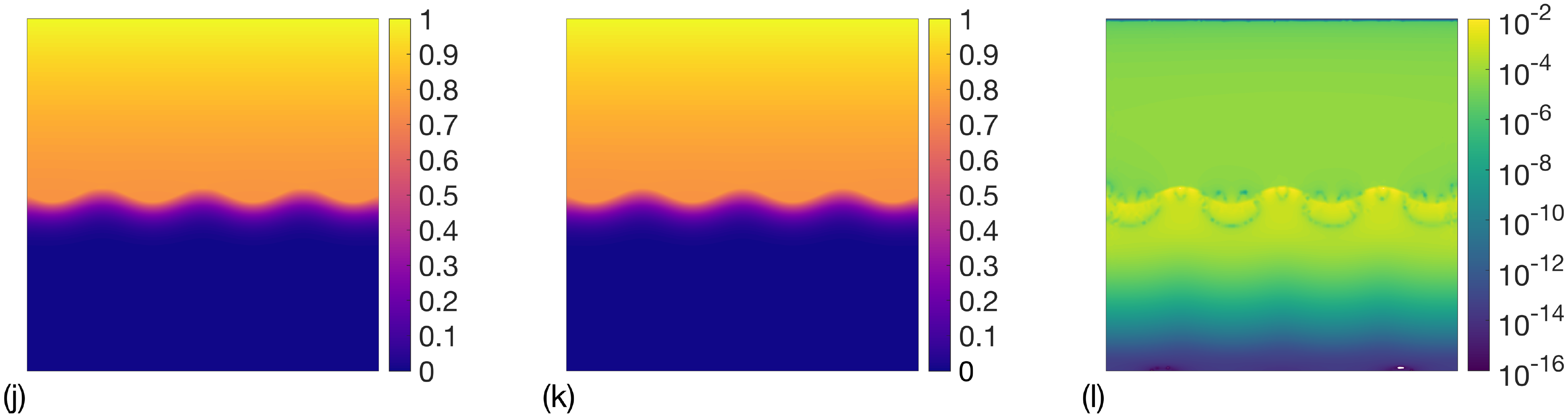}
\caption{Perturbation solution (first column -- (a)(d)(g)(j)), numerical solution (second column -- (b)(e)(h)(k)) and absolute difference between the perturbation and numerical solutions (third column -- (c)(f)(i)(l)) at $t = 0.2$ for four different choices of the interface $x = 0.5 + \varepsilon w(y)$ (a)--(c) $\varepsilon = 0$, $w(y) = 0$, (d)--(f) $\varepsilon = 0.05$, $w(y) = y$, (g)--(i) $\varepsilon = 0.05$, $w(y) = \sin(\pi y)$ and (j)--(k) $\varepsilon = 0.02$, $w(y) = \sin(7\pi y)$. The semi-analytical perturbation solution is computed using the first $30$ terms in the eigenfunction expansions (\ref{eq:U1n})--(\ref{eq:U2n}) and (\ref{eq:diffU1n})--(\ref{eq:diffU2n}) and the first five terms in the perturbation expansions (\ref{eq:u1})--(\ref{eq:u2}). Colormaps available from \cite{cobeldick_2021}.} 
\label{fig:results1}
\end{figure}

Results in Figure \ref{fig:results1} show the comparison between the perturbation and numerical solutions when using the first five terms in the perturbation expansions (\ref{eq:u1})--(\ref{eq:u2}). From these plots, it is evident that the perturbation solution captures the solution behaviour remarkably well and is in good agreement with the numerical solution in all cases. Figures \ref{fig:results1}(a)--(c), show results for the unperturbed problem computed by setting $\varepsilon=w(y) = 0$. These results highlight the best case scenario: \revision{the leading order term in the perturbation solution (section \ref{sec:leading_order}) provides an exact solution to the problem in the Laplace domain and thus any discrepancies between the perturbation and numerical solutions are fully explained by a combination of (i) the approximation error incurred from numerical inversion of the Laplace transforms $u_{1}^{(0)}(x,t) = \mathcal{L}^{-1}\{U_{1}^{(0)}(x,s)\}$ and $u_{2}^{(0)}(x,t) = \mathcal{L}^{-1}\{U_{2}^{(0)}(x,s)\}$ and (ii) the spatial/temporal discretisation error associated with the finite volume method}. Comparing Figures \ref{fig:results1}(f)(i)(l) to Figure \ref{fig:results1}(c) we see that differences between the perturbation and numerical solutions are comparable to those for the unperturbed domain, which is an encouraging sign for the accuracy of the higher order terms, $u_{1}^{(n)}$ and $u_{2}^{(n)}$ ($n = 1,\hdots,4$), in the asymptotic expansions.

\begin{figure}[p]
\centering
\includegraphics[width=0.97\textwidth]{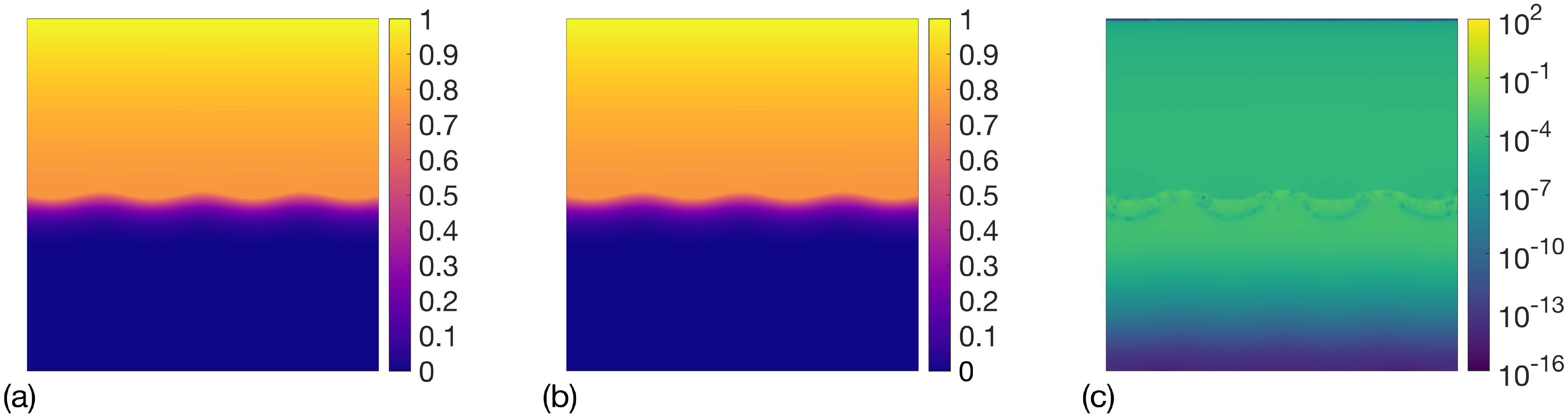}
\includegraphics[width=0.97\textwidth]{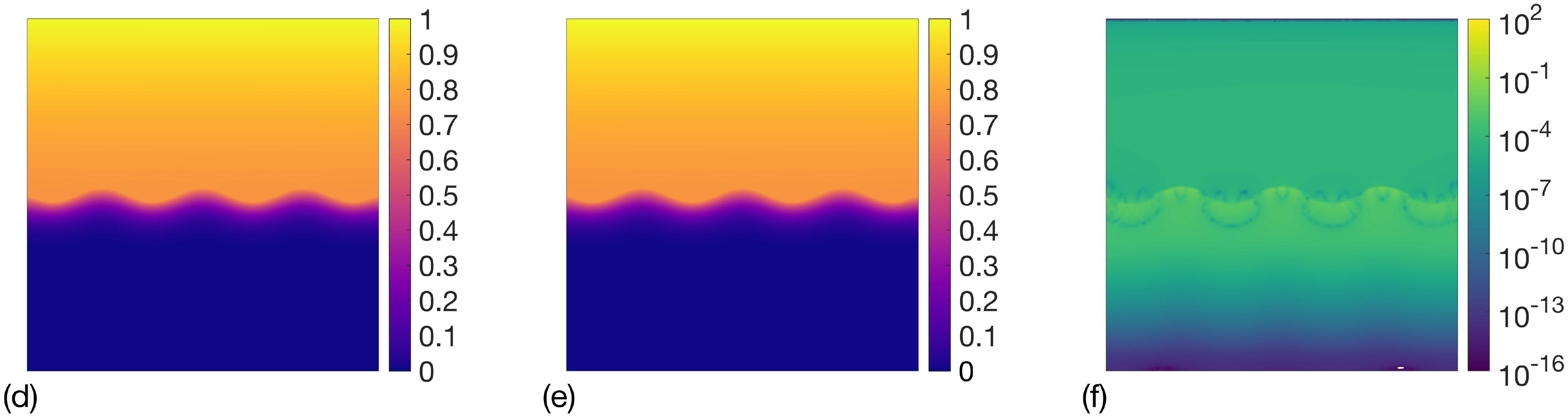}
\includegraphics[width=0.97\textwidth]{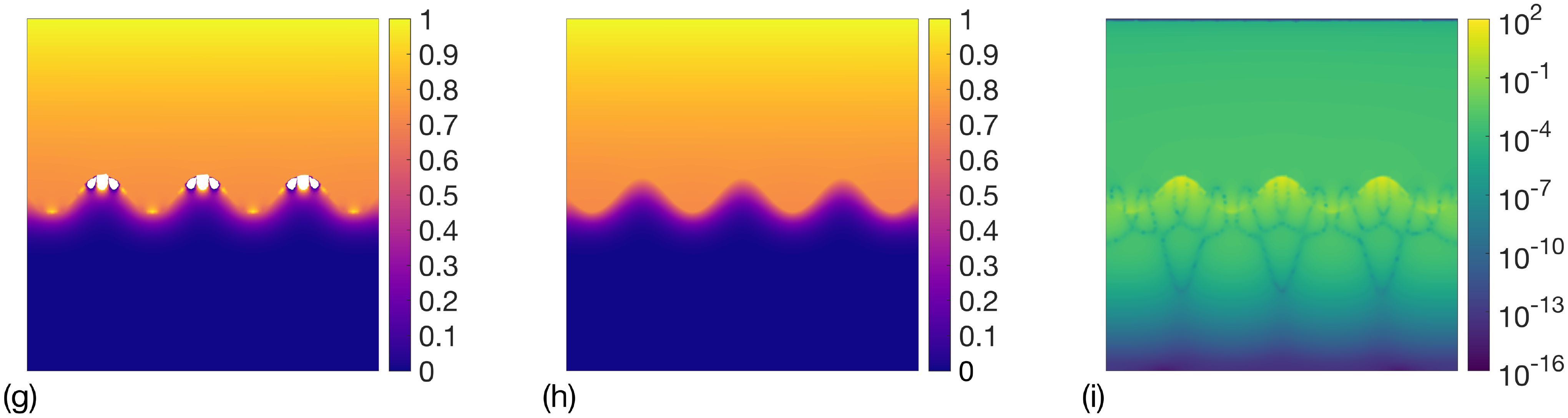}
\caption{Perturbation solution (first column -- (a)(d)(g)), numerical solution (second column -- (b)(e)(h)) and absolute difference between the perturbation and numerical solutions (third column -- (c)(f)(i)) at $t = 0.2$ for $w(y) = \sin(7\pi y)$ and three different choices of the perturbation parameter (a)--(c) $\varepsilon = 0.01$ (d)--(f) $\varepsilon = 0.02$ (g)--(i) $\varepsilon = 0.05$. The semi-analytical perturbation solution is computed using the first $30$ terms in the eigenfunction expansions (\ref{eq:U1n})--(\ref{eq:U2n}) and (\ref{eq:diffU1n})--(\ref{eq:diffU2n}) and the first five terms in the perturbation expansions (\ref{eq:u1})--(\ref{eq:u2}). In (g) white shading indicates regions where the solution falls outside of the range $[0,1]$. Colormaps available from \cite{cobeldick_2021}.} 
\label{fig:results2}
\end{figure}

As is the case for all perturbation methods, the accuracy of our perturbation solution depends on the value of $\varepsilon$ and the number of terms, $N$, taken in the expansions (\ref{eq:u1})--(\ref{eq:u2}). Results in Table \ref{tab:results} provide the maximum absolute difference between the perturbation and numerical solutions (across all vertices in the mesh) for all 45 combinations of $w(y) = y,\sin(\pi y),\sin(7\pi y)$, $\varepsilon = 0.01,0.02,0.05$ and $N = 1,2,3,4,5$ terms in the expansions (\ref{eq:u1})--(\ref{eq:u2}) \revision{with the specific case of $w(y) = \sin(7\pi y)$ and $N = 5$ shown visually in Figure \ref{fig:results2} for $\varepsilon = 0.01,0.02,0.05$.} As expected the match between the perturbation and numerical solutions \revision{improves when $\varepsilon$ is decreased but does not necessarily improve when $N$ is increased, a common feature of perturbation solutions \cite{holmes_2013}.} These results also highlight that care must be taken when using the perturbation solution as it can be unreliable if $\varepsilon$ is too large, as evident in \revision{Figure \ref{fig:results2}(g)--(i) where the accuracy of the solution has deteriorated around the interface. Finally, we} remark also that the perturbation and numerical solutions compared similarly well at later times, for example, a maximum absolute difference of \num{6.97e-03} was recorded at $t = 1$ for $w(y) = \sin(7\pi y)$, $\varepsilon = 0.02$ and $N = 5$ compared with \num{1.13e-02} at $t = 0.2$ (as per Table \ref{tab:results}).

\begin{table}[!t]
\def\arraystretch{0.8}
\centering
\begin{tabular*}{0.9\textwidth}{@{\extracolsep{\fill}}cllll}
$N$ & $\varepsilon$ & $w(y) = y$ & $w(y) = \sin(\pi y)$ & $w(y) = \sin(7\pi y)$\\
\hline
& $0.01$ & \num{1.35e-01} & \num{1.23e-01} & \num{1.62e-01}\\
1 & $0.02$ & \num{2.29e-01} & \num{2.24e-01} & \num{3.35e-01}\\
& $0.05$ &\num{4.90e-01} & \num{4.20e-01} & \num{8.90e-01}\\
\hline
& $0.01$ & \num{9.49e-03} & \num{9.07e-03} & \num{2.96e-02}\\
2 & $0.02$ & \num{3.67e-02} & \num{3.50e-02} & \num{1.32e-01}\\
& $0.05$ & \num{1.97e-01} & \num{1.85e-01} & \num{1.47e-00}\\
\hline
& $0.01$ & \num{9.80e-04} & \num{8.93e-04} & \num{5.83e-03}\\
3 & $0.02$ & \num{2.88e-03} & \num{2.84e-03} & \num{4.03e-02}\\
& $0.05$ & \num{4.36e-02} & \num{4.05e-02} & \num{1.92e+00}\\
\hline
& $0.01$ & \num{7.17e-04} & \num{8.82e-04} & \num{5.41e-03}\\
4 & $0.02$ & \num{1.01e-03} & \num{1.42e-03} & \num{2.17e-02}\\
& $0.05$ & \num{7.22e-03} & \num{6.61e-03} & \num{3.27e+00}\\
\hline
& $0.01$ & \num{7.07e-04} & \num{8.82e-04} & \num{5.43e-03}\\
5 & $0.02$ & \num{8.11e-04} & \num{1.42e-03} & \num{1.13e-02}\\
& $0.05$ & \num{6.48e-03} & \num{3.90e-03} & \num{1.27e+01}\\
\hline
\end{tabular*}
\caption{Maximum absolute difference between the perturbation and numerical solutions (across all vertices in the mesh) at $t = 0.2$ for \revision{different} choices of $\varepsilon$, $w(y)$ and number of terms, $N$, taken in the perturbation expansions (\ref{eq:u1})--(\ref{eq:u2}). In all cases, the perturbation solution is computed using the first $30$ terms in the eigenfunction expansions (\ref{eq:U1n})--(\ref{eq:U2n}) and (\ref{eq:diffU1n})--(\ref{eq:diffU2n}). For reference a maximum absolute difference of \num{6.81e-4} was recorded for the case of $\varepsilon = w(y) = 0$ shown in Figure \ref{fig:results1}(a)--(c). }
\label{tab:results}
\end{table}

\section{Conclusion}
\label{sec:conclusion}
In summary, we have developed a perturbation solution for the problem of time-dependent diffusion across a perturbed interface separating two finite regions of distinct diffusivity. Analogous to similar problems on perturbed domains with constant diffusivity, our analysis shows that each term in the asymptotic expansion satisfies an initial-boundary value problem on the unperturbed domain subject to interface conditions involving the previously determined terms in the asymptotic expansion. Demonstration of the perturbation solution was carried out for a specific, practically-relevant set of initial and boundary conditions and several choices for the perturbed interface with reported results shown to be in good agreement with a standard numerical solution obtained via finite volume discretisation.

The developed solutions expand the suite of solutions for transient diffusion problems and provide analytical insight into important practical problems arising in heat conduction and groundwater contamination. Our perturbation method presented in Sections \ref{sec:solution_method} and \ref{sec:test_case} is quite general allowing for different choices of $\varepsilon$, $w(y)$, $\ell$, $D_{1}$, $D_{2}$, $L$ and $H$ and arbitrary numbers of terms in the asymptotic and eigenfunction expansions. \revision{Although the solutions are semi-analytical due to the application of a numerical inverse Laplace transform, they retain the desirable property of being closed-form expressions that can be evaluated at any point in continuous space and time. This property distinguishes our semi-analytical solutions from standard numerical solutions that discretise the governing equations in space and time and require small temporal and spatial discretisation step sizes to ensure sufficient accuracy.}

The semi-analytical solution presented in Section \ref{sec:test_case} is limited to a specific set of boundary conditions, however, extension to other (perhaps more sophisticated) boundary conditions is fairly straightforward given that the initial-boundary value problem (for each term in the asymptotic expansion) applies on the unperturbed domain. Possible avenues for future work include perturbing the boundary at $x = 0$ or the boundary at $x = L$ instead of the interface, considering a three layer problem with one or two perturbed interfaces or treating more complex governing equations such as diffusion-decay equations.

\section*{Acknowledgements}
We acknowledge funding from Queensland University of Technology's (QUT) Vacation Research Experience Scheme (VRES), which provided DJO with a stipend to undertake this research over the 2020-2021 Australian summer. \revision{We thank Professor Mark McGuinness and one other anonymous referee for their helpful suggestions.}

\appendix
\section{}
\noindent Variables appearing in the solutions $U_{1}^{(0)}(x,s)$ and $U_{2}^{(0)}(x,s)$ (\ref{eq:U10})--(\ref{eq:U20}):
\begin{gather*}
\mu_{1}(s) = \sqrt{s/D_{1}}\,,\quad\mu_{2} = \sqrt{s/D_{2}}\,,\\
\gamma_{11}(s) = -D_{1}\mu_{1}(s)\exp(-\mu_{1}(s)\ell)C_{0}(s)/\psi_{1}(s),\quad \gamma_{12}(s) = -1/\psi_{1}(s),\\
\gamma_{13}(s) = -D_{1}\mu_{1}(s)\exp(\mu_{1}(s)\ell)C_{0}(s)/\psi_{1}(s),\quad \gamma_{14}(s) = 1/\psi_{1}(s),\\
\gamma_{21}(s) = D_{2}\mu_{2}(s)\exp(-\mu_{2}(s)\ell)\revision{Q_{L}}(s)/\psi_{2}(s),\quad \gamma_{22}(s) = -\exp(-\mu_{2}(s)L)/\psi_{2}(s),\\
\gamma_{23}(s) = D_{2}\mu_{2}(s)\exp(\mu_{2}(s)\ell)\revision{Q_{L}}(s)/\psi_{2}(s),\quad \gamma_{24}(s) = -\exp(\mu_{2}(s)L)/\psi_{2}(s),\\
\psi_{1}(s) = -D_{1}\mu_{1}(s)\left[\exp(-\mu_{1}(s)\ell) + \exp(\mu_{1}(s)\ell)\right], \\
\psi_{2}(s) = -D_{2}\mu_{2}(s)\left[\exp(-\mu_{2}(s)(L-\ell)) - \exp(\mu_{2}(s)(L-\ell))\right].
\end{gather*}
\noindent Variables appearing in the solutions $U_{1}^{(n)}(x,y,s)$ and $U_{2}^{(n)}(x,y,s)$ (\ref{eq:U1n})--(\ref{eq:U2n}):
\begin{gather*}
\lambda_{m} = \frac{m\pi}{H},\quad \mu_{1,m}(s) = \sqrt{\frac{s}{D_{1}} + \lambda_{m}^2},\quad \mu_{2,m}(s) = \sqrt{\frac{s}{D_{2}} + \lambda_{m}^2},\\
\widetilde{\alpha}_{m}^{(n)}(s) = \begin{cases} \dfrac{1}{HD_{1}\mu_{1,m}(s)\cosh(\mu_{1,m}(s)\ell)}, & \text{if $m = 0$},\\ \dfrac{2}{HD_{1}\mu_{1,m}(s)\cosh(\mu_{1,m}(s)\ell)}, & \text{if $m\in\mathbb{N}^{+}$},\end{cases}\\
\widetilde{\beta}_{m}^{(n)}(s) = \begin{cases} \dfrac{1}{HD_{2}\mu_{2,m}(s)\sinh(\mu_{2,m}(s)[\ell-L])}, & \text{if $m = 0$},\\ \dfrac{2}{HD_{2}\mu_{2,m}(s)\sinh(\mu_{2,m}(s)[\ell-L])}, & \text{if $m\in\mathbb{N}^{+}$}.
\end{cases}
\end{gather*}
Derivatives appearing in definitions of $F^{(n)}(y,s)$ and $G^{(n)}(y,s)$ (\ref{eq:Fn})--(\ref{eq:Gn}):
\begin{gather}
\nonumber
\frac{\partial^{i+j} U_{1}^{(0)}}{\partial x^{i}\partial y^{j}}(\ell,s) = \frac{1}{2}[1 + (-1)^{j}]\left\{\left[\gamma_{11}(s) + \gamma_{12}(s)V^{(0)}(s)\right]\mu_{1}(s)^{i}\exp(\mu_{1}(s)\ell)\right.\\\label{eq:diffU10}\hspace*{16em} + \left.\left[\gamma_{13}(s) + \gamma_{14}(s)V^{(0)}(s)\right](-\mu_{1}(s))^{i}\exp(-\mu_{1}(s)\ell)\right\}\!,\\
\nonumber
\frac{\partial^{i+j} U_{2}^{(0)}}{\partial x^{i}\partial y^{j}}(\ell,s) = \frac{1}{2}[1 + (-1)^{j}]\left\{\left[\gamma_{21}(s) + \gamma_{22}(s)V^{(0)}(s)\right]\mu_{2}(s)^{i}\exp(\mu_{2}(s)\ell))\right.\\ \label{eq:diffU20}\hspace*{16em} + \left.\left[\gamma_{23}(s) + \gamma_{24}(s)V^{(0)}(s)\right](-\mu_{2}(s))^{i}\exp(-\mu_{2}(s)\ell)\right\}\!,\\
\label{eq:diffU1n}
\frac{\partial^{i+j} U_{1}^{(n)}}{\partial x^{i}\partial y^{j}}(\ell,y,s) = \frac{1}{2}\sum_{m=0}^{\infty} \alpha_{m}^{(n)}(s)\mu_{1,m}(s)^{i}\lambda_{m}^{j}\left[e^{\mu_{1,m}(s)\ell} + (-1)^{i+1}e^{-\mu_{1,m}(s)\ell}\right]\cos(\lambda_{m}y + \frac{j\pi}{2}),\\
\label{eq:diffU2n}
\frac{\partial^{i+j} U_{2}^{(n)}}{\partial x^{i}\partial y^{j}}(\ell,y,s) = \frac{1}{2}\sum_{m=0}^{\infty} \beta_{m}^{(n)}(s)\mu_{2,m}(s)^{i}\lambda_{m}^{j}\left[e^{\mu_{2,m}(s)(\ell-L)} + (-1)^{i}e^{-\mu_{2,m}(s)(\ell-L)}\right]\cos(\lambda_{m}y + \frac{j\pi}{2}),
\end{gather}
for $i \in \mathbb{N}$ and $j \in\{0,1\}$.

\bibliographystyle{elsarticle-num-names}
\bibliography{references}

\end{document}